\documentclass[pra,reprint,preprintnumbers,amsmath,amssymb,superscriptaddress,showpacs,showkeys,longbibliography]{revtex4-2}

\usepackage{graphicx}
\usepackage[normalem]{ulem} 
\usepackage{dcolumn}
\usepackage{bm}
\usepackage{color}
\usepackage{bbold}
\usepackage{braket}
\usepackage{gensymb} 
\usepackage{lipsum}
\usepackage{amsmath}
\usepackage{siunitx}
\usepackage{mathtools} 
\usepackage{extarrows}
\usepackage{natbib}
\usepackage{circledsteps}
\usepackage{array}
\newcolumntype{P}[1]{>{\centering\arraybackslash}p{#1}}
\usepackage{tabularx}

\usepackage[english]{babel}
\usepackage[
	colorlinks=true,
	frenchlinks=false,
        linkcolor=blue,
	    anchorcolor=blue,
        citecolor=blue,
        filecolor=blue,
        urlcolor=blue,
        bookmarks=true,
        bookmarksopen=true,
	    bookmarksnumbered=true,
        bookmarksopenlevel=1,
        plainpages=false,
        pdfpagelabels=true,
	    draft=false]{hyperref}

\DeclareSIUnit\inch{in}
\DeclareSIUnit\bit{bit}
\DeclareSIUnit\dBm{dBm}
\DeclareSIUnit\mol{mol}
\DeclareSIUnit\weber{Wb}
\DeclareSIUnit\sccm{sccm}
\DeclareSIUnit\minute{min}

\addto\captionsenglish{}

\makeatletter
\def\p@subsection{}
\makeatother

\begin{document}

\title{Two-dimensional Planck spectroscopy for microwave photon calibration}

\author{S.~Gandorfer}
\email{simon.gandorfer@wmi.badw.de}
\affiliation{Walther-Mei{\ss}ner-Institut, Bayerische Akademie der Wissenschaften, 85748 Garching, Germany}
\affiliation{Technical University of Munich, TUM School of Natural Sciences, Physics Department, 85748 Garching, Germany}

\author{M.~Renger}
\altaffiliation[S. G. and M. R. contributed equally]{}
\affiliation{Walther-Mei{\ss}ner-Institut, Bayerische Akademie der Wissenschaften, 85748 Garching, Germany}
\affiliation{Technical University of Munich, TUM School of Natural Sciences, Physics Department, 85748 Garching, Germany}

\author{W.~K.~Yam}
\affiliation{Walther-Mei{\ss}ner-Institut, Bayerische Akademie der Wissenschaften, 85748 Garching, Germany}
\affiliation{Technical University of Munich, TUM School of Natural Sciences, Physics Department, 85748 Garching, Germany}

\author{F.~Fesquet}
\affiliation{Walther-Mei{\ss}ner-Institut, Bayerische Akademie der Wissenschaften, 85748 Garching, Germany}
\affiliation{Technical University of Munich, TUM School of Natural Sciences, Physics Department, 85748 Garching, Germany}

\author{A.~Marx}
\affiliation{Walther-Mei{\ss}ner-Institut, Bayerische Akademie der Wissenschaften, 85748 Garching, Germany}

\author{R.~Gross}
\affiliation{Walther-Mei{\ss}ner-Institut, Bayerische Akademie der Wissenschaften, 85748 Garching, Germany}
\affiliation{Technical University of Munich, TUM School of Natural Sciences, Physics Department, 85748 Garching, Germany}
\affiliation{Munich Center for Quantum Science and Technology (MCQST), Schellingstr. 4, 80799 Munich, Germany}

\author{K.\,G.~Fedorov}
\email{kirill.fedorov@wmi.badw.de}
\affiliation{Walther-Mei{\ss}ner-Institut, Bayerische Akademie der Wissenschaften, 85748 Garching, Germany}
\affiliation{Technical University of Munich, TUM School of Natural Sciences, Physics Department, 85748 Garching, Germany}
\affiliation{Munich Center for Quantum Science and Technology (MCQST), Schellingstr. 4, 80799 Munich, Germany}

\date{\today}

\begin{abstract}
Quantum state tomography of weak microwave signals is an important part of many protocols in the field of quantum information processing with superconducting circuits. This step typically relies on an accurate \textit{in-situ} estimation of signal losses in the experimental set-up and requires a careful photon number calibration. Here, we present an improved method for the microwave loss estimation inside of a closed cryogenic system. Our approach is based on Planck's law and makes use of independent temperature sweeps of individual parts of the cryogenic set-up. Using this technique, we can experimentally resolve changes in microwave losses of less than 0.1 dB in the cryogenic environment. We discuss potential applications of this approach for precise characterization of quantum-limited superconducting amplifiers and in other prominent experimental settings.

\end{abstract}

\maketitle

\section{Introduction}
Detection and analysis of microwave quantum signals belong to fundamental tasks in the modern field of superconducting quantum technologies~\cite{Clerk.2010}. The low energy of microwave photons requires cooling to millikelvin temperatures to suppress thermal noise at GHz frequencies and to preserve quantum properties of the related systems. Moreover, in order to experimentally extract information from these quantum signals in the microwave regime, one has to apply sophisticated amplification and detection schemes for quantum state tomography~\cite{Menzel.2012,Eichler.2011,DiCandia.2014}. This state tomography is a crucial building block in all experimental protocols with propagating microwaves or quantum bits, such as entanglement generation~\cite{Menzel.2012,Flurin.2012,Schneider.2020}, quantum state transfer~\cite{Kurpiers.2018,Bienfait.2019,Magnard.2020}, remote state preparation of squeezed states~\cite{Pogorzalek.2019}, quantum teleportation~\cite{Fedorov.2021}, and readout of superconducting quantum circuits~\cite{Blais.2004,Krantz.2019,Schoelkopf.2003}.
 
\begin{figure*}[ht]
	\begin{center}
		\includegraphics[width=2\columnwidth,angle=0,clip]{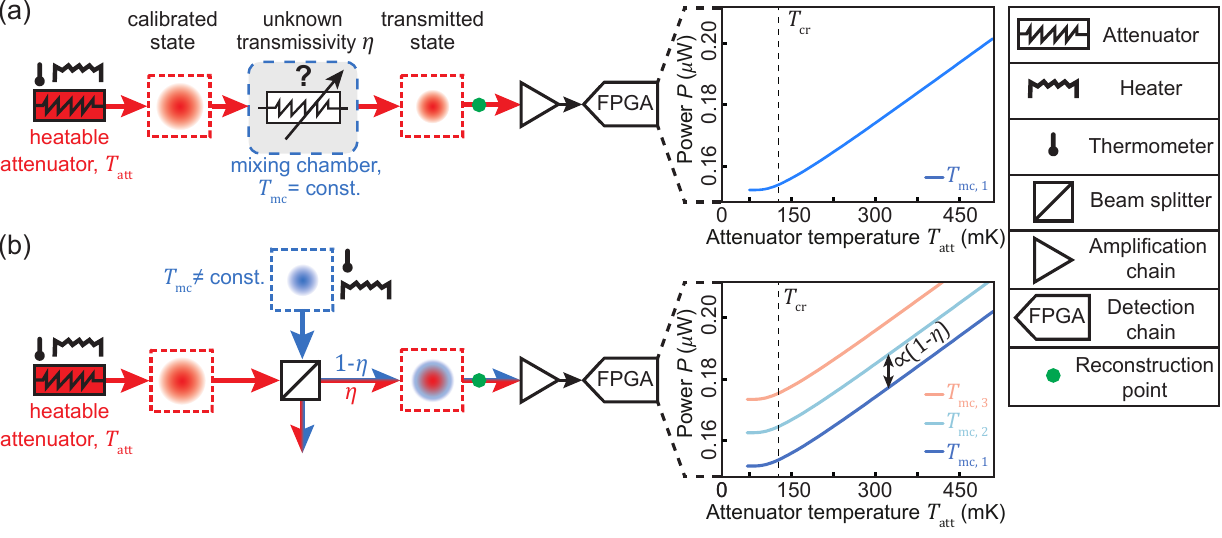}
	\end{center}
	\caption{(a) Schematic drawing of the conventional Planck spectroscopy in a cryogenic system. Thermal radiation emitted by a \SI{30}{dB} heatable attenuator passes through a to-be-calibrated cryogenic set-up consisting of various microwave components with unknown transmissivity, such as superconducting circuits or passive microwave building blocks, before reaching the desired signal reconstruction point (marked by the green dot). In order to correctly perform the photon number calibration, one needs to estimate~$\eta$, which can be done based on datasheet values (unreliable) or by performing the two-dimensional Planck spectroscopy. (b)~Schematic illustration of the two-dimensional Planck spectroscopy. Thermal noise at the respective mixing chamber temperature $T_\text{mc}$ couples to a signal channel via unknown losses,~$1-\eta$. These losses are modelled by an asymmetric beam splitter with transmissivity~$\eta$. Next, the signal is amplified by a cryogenic HEMT amplifier and detected at room temperature using a field programmable gate array (FPGA). The spacing between individual Planck curves, recorded for different $T_{\mathrm{mc,}i}$, allows to extract the unknown transmissivity~$\eta$. The mixing chamber temperatures are chosen such that $T_\text{mc,1} <T_\text{mc,2}<T_\text{mc,3}$. }
	\label{Fig:Setup drawing}
\end{figure*}

A particular challenge for the quantum state tomography in cryogenic experiments is to establish a connection between classical quantities (voltages, currents, etc.) measured at room temperature with conventional receivers to their quantum counterparts (photon numbers, probability amplitudes, etc.) inside of a closed cryogenic set-up. Here, the difficulty lies in the fact that it is impossible to directly access various parts of the experimental set-up during the operation of cryogenic systems at millikelvin temperatures. Therefore, a precise, \textit{in-situ}, characterization procedure of the detection chain is required, providing us with the exact gain and the noise added to a signal by the detection chain. A calibration by sending classical probe signals through the measurement input-output chain is usually not sufficiently precise. This is due to the inaccessibility of an accurate estimation of actual transmission losses of numerous cryogenic microwave components, susceptible to change at cryogenic temperatures~\cite{Pobell.1996,Giordano.1975,Kushino.2008,Kurpiers.2017}. A well established approach to this problem is to embed a reference photon source in the experimental set-up. Many existing calibration techniques rely on quasi black-body radiators emitting well-defined thermal radiation following Planck's law~\cite{Mariantoni.2010,Perelshtein.2022,Simbierowicz.2021,Jebari.2018}, voltage-biased tunnel junctions generating shot noise~\cite{Spietz.2003,Roy.2018}, or qubit-based approaches~\cite{Macklin.2015,Decrinis.2020}. For power calibration of incident microwave signals, similar approaches based on qubits~\cite{Qiu.2023} or bolometer-type detectors~\cite{Girad.2023} can be used. All of the aforementioned calibration techniques are also employed for characterization of the noise performance of cryogenic amplifiers, where a full knowledge of a signal at the amplifier input is needed~\cite{Renger.2021,Simbierowicz.2021,Pozar.2012,Heinz.2022,Bartram.2023}. Here, an accurate characterization of near quantum-limited amplifiers lays the foundation for many experiments that rely on efficient detection of weak microwave signals, such as qubit readout~\cite{Blais.2004,Eddins.2019} or search for dark matter axions~\cite{Bartram.2023,Lasenby.2020}.

Typically, the process of quantum state tomography refers to a specific reconstruction point along the signal path~\cite{Fedorov.2021,Pogorzalek.2019}. This reconstruction point is characterized by the transmissivity value between the reference photon source and the point itself. This nonideal transmissivity,~$\eta <1$, may stem from various microwave components along the signal path which typically introduce finite microwave losses. The precise knowledge of~$\eta$ is crucial for the accurate reconstruction of the quantum states. Therefore, it is of paramount importance to reduce any uncertainty in $\eta$ to a minimum. Often, this goal can be achieved by connecting the calibrated photon source to the device under test by a low-loss superconducting coaxial cable, resulting in $\eta \approx 1$. However, in more complex experiments, this approach may become impractical. There, one has to connect several lossy components following the photon source and rely on an approximate estimation of their microwave losses, based on datasheet values or previous characterization measurements, which are usually carried out at \SI{4}{K} or at room temperature. This approach has been actively used in past complex experiments~\cite{Fedorov.2021,Pogorzalek.2019,Bartram.2023}, which always raises concerns about its impact on experimental results and conclusions.

In this paper we report on an experimental approach aimed at removing the loss estimation step from the quantum state tomography procedure and, thus, achieving a fully self-consistent photon number calibration procedure. Our calibration method is based on the conventional Planck spectroscopy~\cite{Mariantoni.2010}, which is extended by introducing two independent temperature sweeps of different parts of the cryogenic set-up. By doing so, we are able to obtain a reliable photon number calibration of the detection chain while at the same time being able to precisely determine the microwave losses between the black-body radiator and a given reference point. In order to test the accuracy of this method, we employ a flux-tunable superconducting metamaterial and investigate the corresponding change of its transmissivity. Finally, we probe the temperature dependence of the losses introduced by microwave components at cryogenic temperatures.

\section{Methods}
Figure~\ref{Fig:Setup drawing}(a) shows the experimental scheme for the conventional Planck spectroscopy, which is based on a reference black-body radiator as calibrated photon source embedded into the cryogenic system. In our experiments, we use a heatable cryogenic \SI{30}{dB} microwave attenuator as a self-calibrated photon source. The use of a black-body radiator is experimentally advantageous, as it is insensitive to imperfections, requires only a minimal number of additional experimental elements such as a heater, and calibrated thermometer, and can be placed directly in the input signal path without the need for additional microwave switches. As compared to the qubit-based approaches mentioned above, there is no need for an initial characterization of the photon source. For more details of other calibration methods, we refer the reader to Refs.~\cite{Mariantoni.2009,Spietz.2003,Macklin.2015,Girad.2023}. Furthermore, Supplemental Information of Ref.~\cite{Qiu.2023} provides a direct comparison between calibration methods based on tunnel junctions generating shot noise and a qubit-based power calibration implemented in the same experimental setup. Here, both methods provide qualitatively similar calibration results, but with an approximately constant offset in the chain amplification gain. The authors attribute this offset to the insertion loss imposed by additional components necessary to operate the tunnel junction, highlighting the importance of an accurate loss estimation procedure.

According to Planck's law, the microwave power $P_\text{att}$ emitted by the attenuator at a temperature $T_\text{att}$ can be expressed as
\begin{align}\notag
     P_\text{att}(T_\text{att}) = \frac{\langle V^2 \rangle}{Z_0} &=  \int_{f_0-\frac{B}{2}}^{f_0+\frac{B}{2}} \frac{h f }{2}\coth \left( \frac{h f}{2k_\text{B}T_\text{att}} \right) df \\ 
 & \simeq \frac{h f_0 B}{2}\coth \left( \frac{h f_0}{2k_\text{B}T_\text{att}} \right) ,
     \label{eq:noise_power}
 \end{align}
where $V$ is the voltage across the attenuator, $Z_0$ is the characteristic impedance of the microwave circuit, $h$ is the Planck constant, $k_\text{B}$ is the Boltzmann constant, $f_0$ is the carrier frequency, and $B \ll f_0$ is the full detection bandwidth. For high attenuator temperatures, this expression reduces to the classical Johnson-Nyquist relation for the thermal noise power, $ P_\text{att} = \langle V^2 \rangle/Z_0= k_\text{B} T_\text{att} B$. To quantify the onset of this linear regime, we introduce the cross-over temperature $T_\text{cr} = hf_0/2k_\text{B}$~\cite{Mariantoni.2009}. It marks the intersection between the constant asymptotic behavior of Eq.~\eqref{eq:noise_power} in the low temperature limit and the linear asymptotic dependence in the high temperature limit. For a carrier frequency of $f_0 =\SI{5.5}{GHz}$, we have $T_\text{cr} = \SI{132}{mK}$. 

In many experiments, it is essential to reconstruct quantum states at a particular location within the experimental set-up rather than directly at the heatable attenuator output, as illustrated in Fig.~\ref{Fig:Setup drawing}(a). This requires a careful calibration of the detection chain relative to a specific reconstruction point, as states at the reconstruction point are related to those at the heatable attenuator via the power transmissivity $\eta$. The conventional Planck spectroscopy lacks the ability to simultaneously calibrate both the detection chain gain and transmissivity $\eta$. While there exist methods, such as through-reflect-line~(TRL)~\cite{Yeh.2013,Ranzani.2013} or short-open-load-through~(SOLT)~\cite{Simbierowicz.2023}, which can be used to obtain calibrated microwave transmission measurements at millikelvin temperatures, they are usually only suited for the characterization of individual microwave components. As these techniques require their own dedicated experimental setup, these schemes are typically not suited for a straightforward \textit{in-situ} calibration of quantum microwaves experiments which rely on many cryogenic components. Therefore, a common approach is to estimate the transmissivity value from individual datasheets of the respective microwave components, or rely on reflectometry measurements at room temperature to estimate impedance mismatches. However, an accurate prediction of the resulting losses at millikelvin temperatures is generally challenging due to different temperature dependencies of the properties of various components.

In order to experimentally determine the transmissivity~$\eta$, we introduce the two-dimensional Planck spectroscopy by sweeping two independent temperatures, namely the environmental temperature, which is provided by the mixing chamber temperature,~$T_\text{mc}$, of the commercial dilution croystat housing the experiment, and the heatable attenuator temperature,~$T_\text{att}$. The corresponding experimental scheme is shown in Fig.~\ref{Fig:Setup drawing}(b). Here, we consider all lossy microwave components as 4-port devices and describe signal losses using a single, asymmetric beam splitter with the associated power transmissivity~$\eta$. The action of the beam splitter is described using the beam splitter relation $B(\eta,P_1,P_2) = \eta P_1 + (1-\eta)P_2$ for two powers $P_1$ and $P_2$ coupled to the relevant output~\cite{Mariantoni.2009}. Thus, signals incident at the input of this fictitious beam splitter are attenuated and a thermal state, with a mean photon number corresponding to the environmental temperature~$T_\text{mc}$, couples to the microwave channel. Respectively, the microwave power measured at room temperature at the end of the amplification chain is given by
\begin{align}\notag
     P &=  \frac{\kappa}{Z_0}\bigg[ \frac{\eta}{2}\coth \left( \frac{h f_0}{2 k_\text{B}T_\text{att}} \right) \\ \label{eq:output_power}
    &+\frac{1-\eta}{2}\coth \left( \frac{h f_0}{2 k_\text{B}T_\text{mc}} \right) +n_\text{H} \bigg] \enspace , 
\end{align}
where $\kappa$ is the photon number conversion factor~(PNCF). This proportionality factor relates the power measured at room temperature to the photon number at the desired reference point in the cryogenic set-up and depends on the bandwidth, the carrier frequency, and the amplification gain. Furthermore, $n_\text{H}$ is the number of noise photons added by our amplification chain, stemming to a large extent from a cryogenic high-electron mobility transistor~(HEMT) amplifier during initial amplification of the signal. Due to typically high gain values of the microwave HEMT amplifiers ($G_\text{H} > \SI{30}{dB}$), we can neglect the noise contribution from the cascaded room temperature amplifiers~\cite{Friis.1944}. The quantity $\kappa / Z_0$ can be interpreted as the detected power per microwave photon. 

\begin{figure*}
	\begin{center}
		\includegraphics[width=2\columnwidth,angle=0,clip]{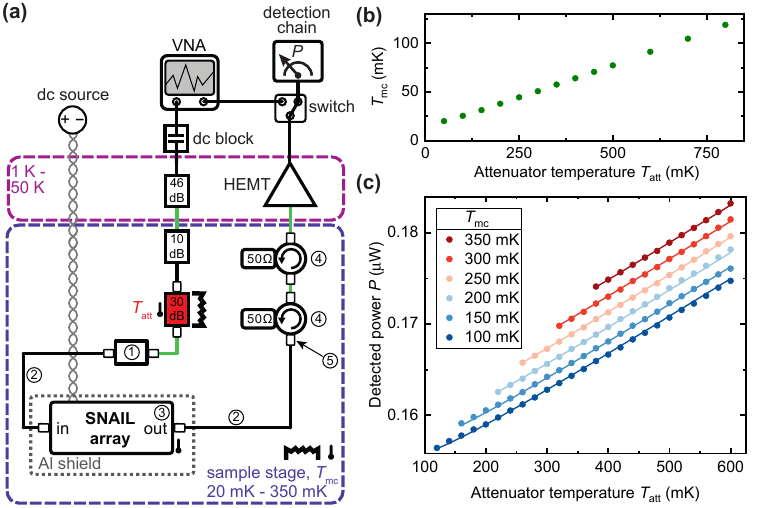}
	\end{center}
	\caption{(a) Experimental set-up for the two-dimensional Planck spectroscopy. Encircled numbers label different, lossy, microwave components placed along the propagation path of a thermal signal emitted by the heatable attenuator. From the respective datasheets, we extract: \Circled{1}\,: directional coupler, CPL4000-8000-20-C, Miteq, \SI{0.16}{dB} of insertion losses, \Circled{2}\,: semi-rigid tin-plated copper coaxial cable, SF-085NM, Isotech, \SI{0.34}{dB} of insertion losses, \Circled{3}\,: SNAIL metamaterial, \SI{0.60}{dB} of insertion losses~\protect\cite{Perelshtein.2022}, \Circled{4}\,: isolator, CTH1 184-KS18, Quinstar \SI{0.17}{dB} of insertion losses, \Circled{5}\,: SMA/SMP microwave connector, \SI{0.05}{dB} of insertation losses each. Overall, this amounts to total estimated (datasheet) losses of \SI{2.18}{dB}. Green and black connections indicate superconducting and normal conducting coaxial microwave cables, respectively. We only indicate SMA/SMP connectors at positions relevant for the photon calibration. (b) Steady-state mixing chamber temperature, $T_\text{mc}$, as a function of the stabilized attenuator temperature, $T_\text{att}$. (c) The experimental two-dimensional Planck spectroscopy of our microwave channel. Symbols depict experimental data and solid lines are corresponding fits according to Eq.~\eqref{eq:output_power}. The losses extracted by using this procedure, $L_\text{2D}$ = \SI{2.79}{dB}, are higher than expected from the datasheet values.}
	\label{Fig:PNCF_0_uA}
\end{figure*}

Analysis of the structure of Eq.~\eqref{eq:output_power} immediately provides an intuition for the working principle of the two-dimensional Planck spectroscopy. We first analyze the power difference, $\Delta P$, measured at room temperature at the end of the amplification chain for two different temperature values of the mixing chamber, $T_\text{mc} = T_\text{mc,1}, T_\text{mc,2}$, while we stabilize the heatable attenuator at a constant temperature. For environment temperatures well above the cross-over temperature $T_{\mathrm{cr}}$, the noise power can be substituted in good approximation with the linear Johnson-Nyquist dependence. In this case, we find
\begin{align} \notag
    \Delta P &=  \left( 1-\eta\right) \left[ P\left(T_\text{mc,2}\right)-P\left(T_\text{mc,1}\right)  \right]  \\ &\simeq  \frac{\kappa}{Z_0}\left( 1-\eta\right) \frac{k_\text{B} \Delta T}{h f_0 } \enspace ,
    \label{Eq:2D_PNCF_spacing}
\end{align}
where $\Delta T \equiv T_\text{mc,2}-T_\text{mc,1}$. Equation~\eqref{Eq:2D_PNCF_spacing} illustrates the fact that we can accurately obtain the unknown transmissivity~$\eta$ by measuring the vertical spacing between different conventional Planck spectroscopy curves shown in Fig.~\ref{Fig:Setup drawing}(b). By extending this method to, at least, three different mixing chamber temperatures we can additionally investigate a potential temperature dependence of~$\eta$.
\section{Experimental results}
\subsection{Experimental set-up}
Our experimental set-up consists of a flux-tunable superconducting circuit serially connected to a cryogenic HEMT amplifier (Low Noise Factory LNF-LNC4\_8C). The superconducting circuit is made from a coplanar transmission line with an array of SNAILs (Superconducting Nonlinear Asymmetric Inductive eLement) which forms a flux-tunable microwave waveguide~\cite{Frattini.2017,Perelshtein.2022}. We use this system to vary the overall attenuation through the microwave channel. At the carrier frequency of \SI{5.5}{GHz}, the metamaterial introduces around \SI{0.6}{dB} of internal losses~\cite{Perelshtein.2022}. 

A detailed scheme of the experimental set-up is shown in Fig.~\ref{Fig:PNCF_0_uA}(a). Except for the HEMT amplifier, all of the cryogenic microwave components are thermally coupled to the mixing chamber plate of the dilution cryostat using high purity, annealed silver wires. The heatable \SI{30}{dB} attenuator is located at the input of the sample stage. A \SI{100}{\ohm} resistor, serving as a heater, and a RuO$_2$ thermometer attached to this attenuator allow for stabilization of its temperature using a PID control loop realized with an external Picowatt AVS-48SI Picobridge resistance bridge. We use a stainless steel input cable and a superconducting NbTi/NbTi coaxial cable after the heatable attenuator. At a temperature of \SI{200}{mK}, the thermal conductivity $\lambda$ of NbTi and stainless steel are $\lambda_\text{NbTi} = \SI{1.08e-5}{ \watt \per \centi \metre \per \kelvin}$~\cite{Olson.1993} and $\lambda_\text{SS} = \SI{1.82e-4}{ \watt \per \centi \metre \per \kelvin}$~\cite{Pobell.1996}, respectively. Both coaxial cables are based on polytetrafluoroethylene (PTFE) as the dielectric material. In this temperature regime, the contribution to the thermal conductivity stemming from the PTFE amounts to $\lambda_\text{PTFE} = \SI{1.2e-6}{ \watt \per \centi \metre \per \kelvin}$~\cite{Drobizhev.2017,Kushino.2008}. Due to the low overall thermal conductivity of both coaxial microwave cables, we use an additional thinned silver ribbon to weakly couple the heatable attenuator to the mixing chamber plate. This allows us to balance a heat load applied to the dilution fridge during temperature sweeps. The thermal conductivity of this silver ribbon sensitively depends on the purity of the material and the annealing process~\cite{Pobell.1996} and we conservatively estimate $\lambda_\text{Ag} = \SI{2e-2}{ \watt \per \centi \metre \per \kelvin}$. Consequently, the thermalization of the heatable attenuator is determined by the heat conductivity $\lambda_\text{Ag}$ of the silver ribbon. The temperature of the mixing chamber plate is stabilized using a second PID control loop based on a Lakeshore Model 372 AC Resistance Bridge. A thermometer mounted to the sample holder containing the SNAIL metamaterial allows us to verify a homogeneous temperature distribution between the mixing chamber plate and the sample holder. We use a heterodyne microwave receiver at room temperature as well as a subsequent digital data processing set-up, similar to the one described in Ref.~\cite{Fedorov.2021}. Here, the signal is filtered around the carrier frequency of \SI{5.5}{GHz} and down-converted to an intermediate frequency (IF) of \SI{12.5}{MHz} using an image rejection mixer. Before and after the down-conversion process, the signal is amplified with multiple amplifiers at room temperature. The signal is digitized with an analog-to-digital converter, digitally down-converted and filtered within a full bandwidth of $B = \SI{400}{kHz}$ around the desired carrier frequency using a finite-impulse response~(FIR) filter. For the last step, the second order quadrature moments~$\langle I^2\rangle$ and~$\langle Q^2 \rangle$, quantifying the signal power, are calculated. All of the aforementioned digital data processing steps are performed with an FPGA. 
\subsection{Measurement results}
First, we measure the dependence of~$T_\text{mc}$ on the attenuator temperature ~$T_\text{att}$. Here, we sweep the temperature of the \SI{30}{dB} heatable attenuator,~$T_\text{att}$, and measure the steady-state temperatures of the mixing chamber plate,~$T_\text{mc}$. The result of this measurement is shown in Fig.~\ref{Fig:PNCF_0_uA}(b). We conclude that the attenuator can only be heated up to \SI{600}{\milli \kelvin}, as long as we want to stabilize the mixing chamber plate at \SI{100}{\milli \kelvin}.  For the two-dimensional Planck spectroscopy we perform six consecutive sweeps of the attenuator temperature, while increasing the mixing chamber temperature in steps of \SI{50}{\milli \kelvin}, starting from $T_\text{mc}=\SI{100}{\milli \kelvin}$. For each step, the lowest heatable attenuator temperature is set 20-\SI{30}{\milli \kelvin} higher than the present mixing chamber temperature to ensure sufficient temperature stabilization within $\pm \SI{0.1}{mK}$ of the desired value. Figure~\ref{Fig:PNCF_0_uA}(c) shows the resulting experimental data. All six experimental Planck curves are simultaneously fitted with a weighted least-square fit, according to Eq.~(\ref{eq:output_power}). The weights for each Planck curve are chosen inversely proportional to the respective number of data points. Furthermore, the aforementioned calibration procedure enables us to treat $T_\text{att}$ and $T_\text{mc}$ as independent control quantities. The measured data shows an approximately equidistant spacing between the different mixing chamber temperatures, as expected from~Eq.~(\ref{Eq:2D_PNCF_spacing}). This indicates a constant transmissivity $\eta$ in the temperature range up to~$\simeq \SI{300}{mK}$. For this fit, we employ $\kappa$, $n_\text{H}$, and $\eta$ as fit parameters and observe a good agreement between the measured and fitted data for $\kappa_\text{2D} = \SI{1.15}{(mV})^2$, $n_\text{H,2D} = 6.83$, and microwave losses $L_\text{2D} = -10\log_{10}\eta =\SI{2.79}{dB}$. The number of HEMT amplifier noise photons measured with the two-dimensional Planck spectroscopy is in good agreement with the datasheet value of $n_\text{H} = 6.20$. At the carrier frequency of \SI{5.5}{GHz}, we can also estimate the signal losses based on the datasheet values of individual components. The contributions of respective microwave components are listed in the caption of Fig.~\ref{Fig:PNCF_0_uA}(a) and we obtain $L_\text{datasheet} = \SI{2.18}{dB}$. We observe that the path loss obtained with the two-dimensional Planck spectroscopy is roughly \SI{0.6}{dB} larger than the value estimated from the datasheet numbers.

\begin{figure*}
	\begin{center}
		\includegraphics[width=2\columnwidth,angle=0,clip]{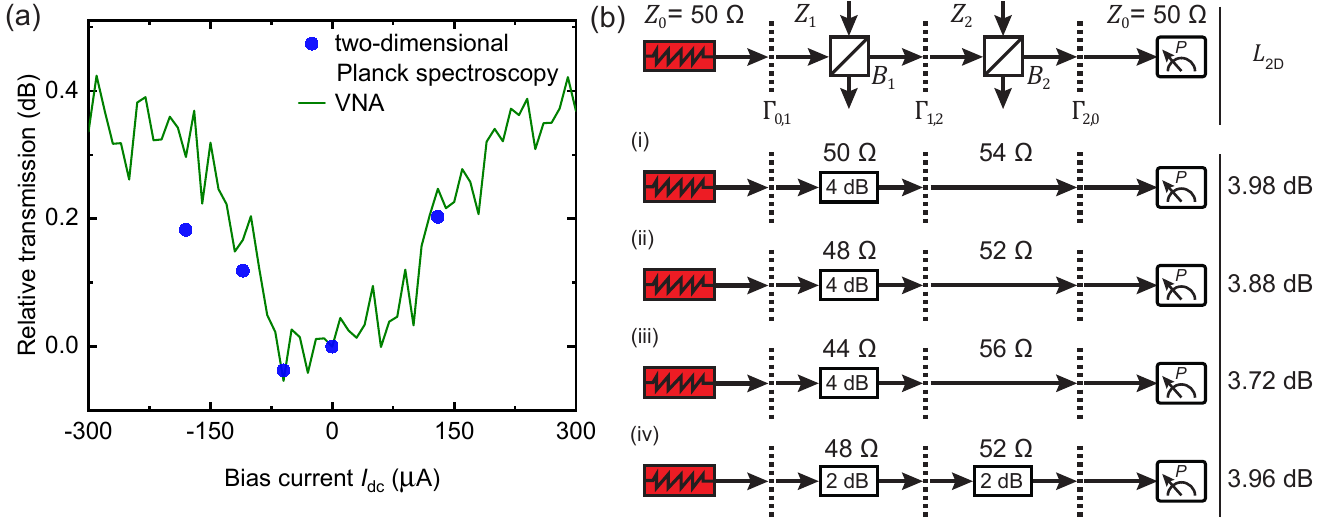}
	\end{center}
	\caption{(a) Relative microwave transmission through the signal path as a function of the bias current through the SNAIL metamaterial measured with a VNA (solid green line) and extracted from the two-dimensional Planck spectroscopy (blue dots). (b) Power losses extracted with the two-dimensional Planck fitting routine under the assumption of different impedance mismatch scenarios (i-iv). The heatable attenuator, depicted in red, is \SI{50}{\Omega} impedance matched. The losses are distributed between two regions with impedances $Z_1$ and $Z_2$ and are modelled via asymmetric beam splitters $B_{1,2}$ with respective transmissivities $\eta_{1,2}$.}
	\label{Fig:PNCF_result}
\end{figure*}

The PNCF $\kappa$ is a fundamental quantity for reconstruction of quantum states~\cite{Renger.2021,Fedorov.2016,Fedorov.2021,Pogorzalek.2019} and slight deviations in its value might have a drastic impact on the properties of the reconstructed quantum states. In order to investigate the effect of underestimating losses, as it is the case when using the values provided by the datasheets, we can compare the PNCF obtained from the two-dimensional Planck spectroscopy,~$\kappa_\text{2D}$, to the one derived from a conventional, one-dimensional, Planck spectroscopy,~$\kappa_\text{1D}$. To this end, we can evaluate the data measured at the fixed mixing chamber temperature of \SI{100}{mK} and use the fitting routine according to Eq.~(\ref{eq:output_power}), but with a fixed bath temperature, corresponding to the conventional Planck calibration measurement. In this way, we do not obtain any information about the unknown transmissivity, $\eta$, present in the propagation path of the signal and have to rely on the datasheet value of $L_\text{datasheet} = \SI{2.18}{dB}$. This underestimation of losses, as compared to the two-dimensional Planck spectroscopy, directly leads to an overestimation of both $\kappa_\text{1D}$ and $n_\text{H}$. Specifically, we obtain values of $\kappa_\text{1D} = \SI{1.34}{(mV})^2$ and $n_\text{H,1D} = 9.59$. Here, it is important to note that we only measure an effective noise photon number which is scaled by the inverse transmissivity.

Now, we can consider a hypothetical reconstruction experiment of a squeezed vacuum state, a quantum state commonly used in many experiments~\cite{Fedorov.2016,Fedorov.2021, Malnou.2018,Eichler.2011,Mallet.2011,Flurin.2012}, based on the PNCF values obtained with both methods. Quantum state parameters, which are often used to describe the properties of these states, are the squeezing level $S$, defined as a suppression of the squeezed signal quadrature variance $(\Delta Q_\mathrm{s})^2$ below the vacuum level, $S = -10 \log_{10}\left[(\Delta Q_\mathrm{s})^2/0.25\right]$~\cite{Zhong.2013}, and the purity $\mu$, which quantifies deviation from the Heisenberg limit and thus an amount of classical noise (mixedness) in the reconstructed quantum state~\cite{Fedorov.2021}. Assuming the hypothetical squeezed state with a squeezing level of $S_\text{2D} = \SI{3.00}{dB}$ and purity of $\mu = 1.00$ obtained with the help of the two-dimensional Planck spectroscopy, the same state will be reconstructed with the squeezing level $S_\text{1D} = \SI{2.45}{dB}$ and purity of $\mu =0.96$, when using the the PNCF from the conventional Planck spectroscopy. This highlights the difference between the two calibration methods as well as the importance of precise knowledge of both the PNCF and losses present in the system for the precise quantum state tomography.

We can further explore the potential of \textit{in-situ} loss measurement via the two-dimensional Planck spectroscopy by utilizing the Josephson metamaterial as a flux-tunable attenuator. To this end, we vary the SNAIL flux bias~$\Phi$ by changing the current,~$I_\text{dc}$, through the on-chip bias line. First, we measure the transmitted power ~$\tau\left(I_\text{dc} \right)$ through the system using a vector network analyzer (VNA). Such measurements cannot be used to estimate the absolute losses in individual subsystems, but provide useful information about a  relative change of losses. To this end, we start by performing an amplitude reference measurement at zero flux bias and the carrier frequency of \SI{5.5}{GHz}, $\tau_\text{ref} = \tau\left(I_\text{dc} = 0 \right)$, with an IF bandwidth of \SI{10}{Hz}. Then, we calculate the change of losses as a function of bias current, $\Delta \tau\left(I_\text{dc} \right) = \tau\left(I_\text{dc} \right)-\tau_\text{ref}$. Subsequently, we perform multiple two-dimensional Planck measurements at different flux biases and use the fitting routine according to Eq.~(\ref{eq:output_power}) to extract the loss values $L\left( I_\text{dc} \right)$. As before, we take the result at zero flux bias, $L\left(I_\text{dc} = 0 \right) = L_\text{ref}$ as a reference and only consider the change of loss with respect to this value. Figure~\ref{Fig:PNCF_result}(a) shows the result of this procedure. We observe good agreement between both methods. The two-dimensional Planck spectroscopy is able to resolve loss changes smaller than \SI{0.1}{dB} in our experimental set-up. For SNAIL bias currents below $I_\text{dc} = \SI{-140}{\mu A}$, the change in attenuation is smaller than expected from the VNA measurements. This deviation may be a result of substantially different measurement bandwidths used in the two methods: while the VNA probes a narrow bandwidth of \SI{10}{Hz}, the two-dimensional Planck spectroscopy bandwidth is defined by a digital FPGA filter with a bandwidth of \SI{400}{kHz}. Respectively, many microwave components, such as circulators or the SNAIL metamaterial, may have a weakly frequency-dependent power loss spectrum~\cite{Perelshtein.2022}. Consequently, this frequency dispersion may lead to deviations of estimated losses when using measurements with different effective bandwidths. 

So far, we have assumed an ideal $\SI{50}{\Omega}$ impedance of all circuit components when performing the two-dimensional Planck-spectroscopy. However, in real experiments, additional reflections in microwave lines can occur due to a finite impedance mismatch between different circuit parts. In order to evaluate effects of these mismatches we model several scenarios shown in Fig.~\ref{Fig:PNCF_result}(b) and display corresponding hypothetical loss values derived from the two-dimensional Planck-spectroscopy. At a circuit position where the circuit impedance changes from $Z_i$ to $Z_j$, a fraction of the signal is reflected according to the reflection coefficient $|\Gamma_{i,j}|^2 = |V_j/V_i|^2= (Z_j - Z_i)^2 /(Z_j + Z_i)^2$~\cite{Pozar.2012}, where $V_j$ is the voltage of the reflected  wave and $V_i$ is the voltage of the incident wave. By tracing the signal voltages throughout the setup, we generate four sets of data according to different characteristic situations illustrated in Fig.~\ref{Fig:PNCF_result}(b). In our analysis, we consider the initial microwave power emitted by heatable attenuator, $P_\text{in} = \langle V_\text{in}^2\rangle/Z_0$, according to Eq.~\ref{eq:noise_power}, placed in a perfect $\SI{50}{\Omega}$ environment. Using a cascaded circuit model, we obtain the final microwave power as
\begin{gather}
\notag
P_\text{out} = \frac{\langle V_\text{out}^2\rangle}{Z_0} = \left(1- \Gamma_{2,0} \right) \left[ \left( 1-\eta_2 \right) \frac{\langle V_\text{mc}^2\rangle}{Z_2} + \right. \\  \left.
\eta_2 \left( 1- \Gamma_{1,2} \right) \left(  \left(1- \eta_1 \right) \frac{\langle V_\text{mc}^2\rangle}{Z_1} + \eta_1 \left( 1-\Gamma_{0,1} \right) \frac{\langle V_\text{out}^2\rangle}{Z_0}  \right) \right]\: .
\end{gather}
Here, as an illustrative example the total power losses are split between two regions with different characteristic impedances. The individual loss components are modelled via asymmetric beam splitter relations, $B_{1,2}$, coupling the incident powers to a noise signal power $\langle V_\text{mc}^2\rangle/Z_{1,2}$ generated by the thermal bath at the mixing chamber temperature. The losses, $L_{2D}$, are extracted by fitting the corresponding datasets with the weighted least-square routine, identical to the one we use to analyze the experimental data. Here, we do not consider any interference effects arising from back-propagating waves due to the sub-nanosecond, i.e., negligibly short, coherence times of thermal fields for temperatures in the range of tens of millikelvin~\cite{Mehta.1963,Kano.1962}. By considering case~(i) in Fig.~\ref{Fig:PNCF_result}(b), we observe that reflections due to finite impedance mismatches have a weak influence on the extracted losses, leading to a reduction of extracted losses to $L_\text{2D}=$~\SI{3.98}{dB} compared to the \SI{4}{dB} assumed for generating the data sets. By comparing the cases~(i)-(iv), we observe that a deviation from the assumed \SI{4}{dB} of internal losses arises from the thermal noise coupled to the signal path via the beam splitter relation, renormalized by the $1/Z_i$ term in Eq.~(\ref{eq:noise_power}). We note that there are situations, where several impedance mismatches can cancel out, as is illustrated by case~(iv) in Fig.~\ref{Fig:PNCF_result}(b). We also investigate the effect of impedance mismatch to the reconstruction of a hypothetical propagating squeezed state with the squeezing level of $S = \SI{3.00}{dB}$. In this case, we obtain a maximal squeezing level deviation for the case~(iii), which corresponds to a decrease to $S = \SI{2.85}{dB}$.

\begin{figure}
	\begin{center}
		\includegraphics[width=\columnwidth,angle=0,clip]{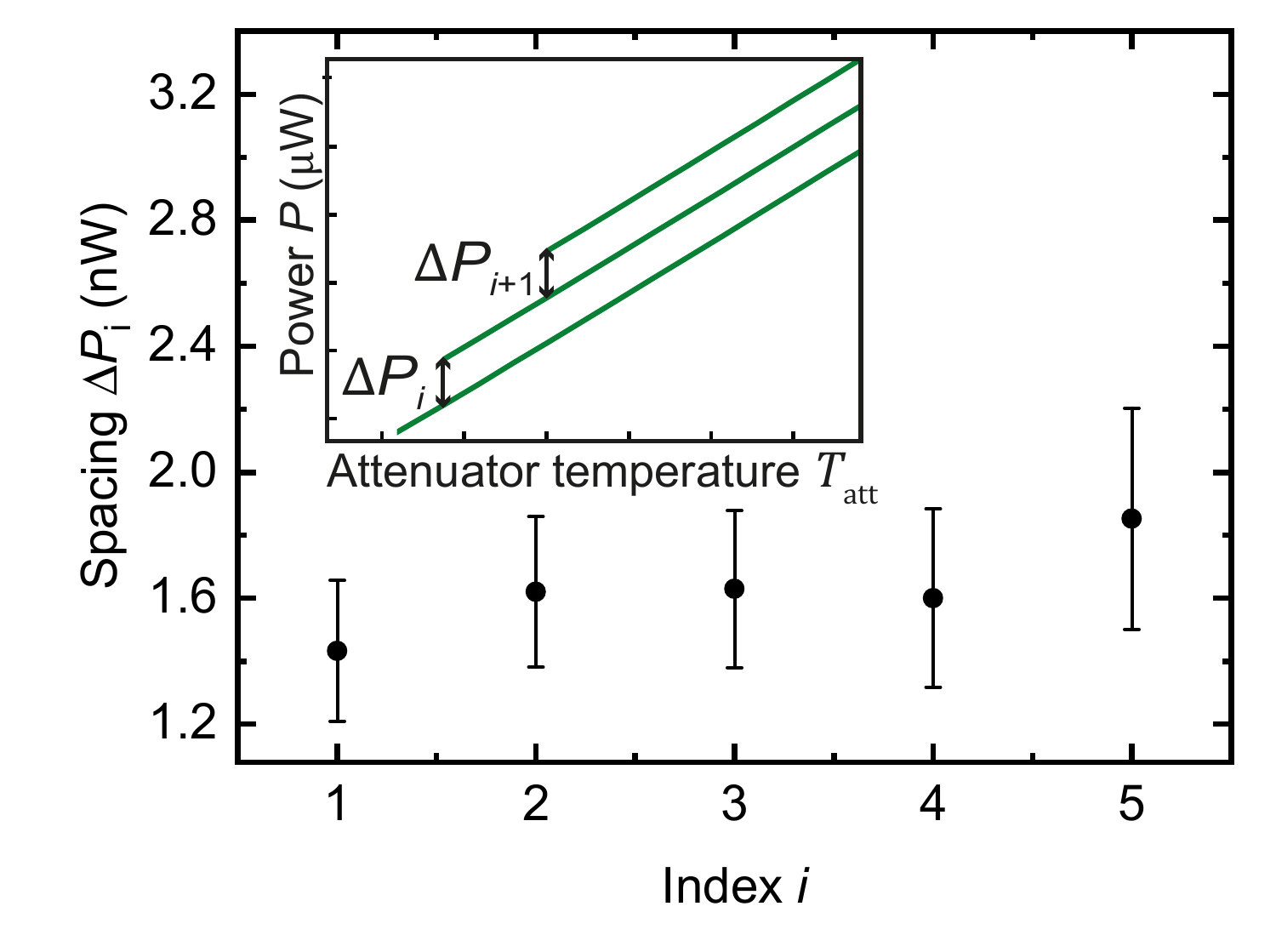}
	\end{center}
	\caption{Spacing $\Delta P_i$ between the Planck curves measured at two adjacent mixing chamber temperatures, as indicated by the inset. The spacing values are evaluated from the measurement dataset shown in Fig.~\ref{Fig:PNCF_0_uA}(c). We see a slightly increasing spacing for the curves recorded at the higher mixing chamber temperatures. }
	\label{Fig:PNCF_spacing}
\end{figure}

Next, we take a closer look at the temperature dependence of the channel transmissivity,~$\eta$, obtained with the two-dimensional Planck spectroscopy. Figure~\ref{Fig:PNCF_spacing} shows the spacing $\Delta P_i$ between the Planck curves measured at two adjacent mixing chamber temperatures, extracted from Fig.~\ref{Fig:PNCF_0_uA}(c). The index $i$ corresponds to the spacing between Planck curves measured at mixing chamber temperatures of $T_\text{mc} = (i+1)\cdot\SI{50}{mK}$ and $T_\text{mc} = (i+2) \cdot\SI{50}{mK}$. According to Eq.(\ref{Eq:2D_PNCF_spacing}), this spacing allows to investigate a temperature dependence of the channel transmissivity. We observe that the curves corresponding to the highest mixing chamber plate temperatures, $T_\text{mc} = \SI{300}{mK}$ and $T_\text{mc} = \SI{350}{mK}$, show a spacing~$\Delta P_5$ that is slightly increased in comparison to the lower temperatures. Consequently, these results indicate an approximately constant transmissivity up to a mixing chamber temperature of \SI{300}{mK}, followed by a subsequent decrease in transmissivity. We attribute this decrease in transmissivity to the temperature-dependent resistance in normal conducting materials in cryogenic circulators~\cite{Allen.1956} or to a possible formation of quasiparticles in the superconducting sample~\cite{Aumentado.2004,Wilson.2004}. We note that the two-dimensional Planck spectroscopy enables the distinction between temperature-dependent and temperature-independent losses. Furthermore, we can quantify this temperature dependence and take it into account for the reconstruction of quantum states at elevated temperatures. For future experiments, one could also apply the two-dimensional Planck spectroscopy as a tool to investigate a genuine noise temperature of a device, as for many systems, an actual electronic temperature may differ from a phononic temperature. This effect is related to microwave dissipation and excitation of the so-called ‘hot electrons’ and the weak coupling of the electronic to the phononic system at low temperatures~\cite{Roukes.1985,Wellstood.1994,Yeh.2019}.

\section{Conclusion}
We have demonstrated an experimental implementation of the two-dimensional Planck spectroscopy allowing for an \textit{in-situ} calibration of microwave losses, as well as for an accurate estimation of photon conversion coefficients in a cryogenic set-up. To this end, we have extended the conventional Planck spectroscopy approach~\cite{Mariantoni.2010} by adding the second degree of freedom in the form of variable temperature of the whole cryogenic environment (the mixing chamber temperature). We have employed a beam splitter loss model to describe and extract an unknown transmissivity (losses) from the two-dimensional Planck spectroscopy measurements, and thus, improve the calibration accuracy. Furthermore, our approach has enabled us to directly assess the temperature dependence of microwave losses. In our experiments, we have observed a good agreement between the measured data and fitted theory model. Moreover, we have employed a Josephson metamaterial sample as a flux-tunable attenuator to benchmark the accuracy of the two-dimensional Planck spectroscopy and demonstrated that we are able to resolve microwave loss changes down to \SI{0.1}{dB}. 

Our calibration technique relies on using a heatable attenuator as a broadband black-body emitter. This source can be employed in a large variety of modern cryogenic experiments, including characterization of cryogenic HEMT amplifiers, quantum-limited parametric amplifiers, and other superconducting devices. As compared to other currently available calibration techniques in microwave cryogenic experiments, the two-dimensional Planck spectroscopy has the specific advantage of being applicable in a very wide frequency range without extra experimental efforts due to the nature of black-body radiation. This black-body radiator can be embedded directly into the respective microwave lines, without a need for dedicated cryogenic microwave switches. Last but not least, the two-dimensional Planck spectroscopy completely relies on off-the-shelf microwave components and does not require fabrication of state-of-the-art superconducting devices. Our analysis also indicates that the two-dimensional Planck spectroscopy is partially resilient to circuit imperfections such as impedance mismatches. Its inherent ability to determine internal signal losses allows for a more precise estimation of the noise performance of the aforementioned devices. We envision important applications of this method in experiments that rely on the efficient detection of signal amplitudes, such as the search for dark matter axions~\cite{Bartram.2023}, where the precise knowledge of noise figures is a key prerequisite. Furthermore, it can also serve as an important tool for characterization of various qubit readout schemes.

\section{Acknowledgments}
We acknowledge support by the German Research Foundation via Germany’s Excellence Strategy (EXC2111-390814868), the German Federal Ministry of Education and Research via the project QUARATE (Grant No.13N15380). This research is part of the Munich Quantum Valley, which is supported by the Bavarian state government with funds from the Hightech Agenda Bayern Plus. We acknowledge Joonas Govenius and VTT Technical Research Centre of Finland Ltd. for providing the SNAIL-based superconducting metamaterial sample.

\bibliography{Bibliography}
\medskip\noindent

\clearpage

\end{document}